\documentclass[%
 aip,
 amsmath,amssymb,
reprint,%
]{revtex4-1}

\usepackage{graphicx}
\usepackage{dcolumn}
\usepackage{bm}

\usepackage[utf8]{inputenc}
\usepackage[T1]{fontenc}
\usepackage{mathptmx}
\usepackage{etoolbox}

\usepackage[hidelinks]{hyperref}
\usepackage[capitalise]{cleveref}
\usepackage{physics}
\usepackage{amssymb}

\DeclareMathOperator{\sinc}{sinc}

\makeatletter
\def\@email#1#2{%
 \endgroup
 \patchcmd{\titleblock@produce}
  {\frontmatter@RRAPformat}
  {\frontmatter@RRAPformat{\produce@RRAP{*#1\href{mailto:#2}{#2}}}\frontmatter@RRAPformat}
  {}{}
}%
\makeatother
\begin{document}

\preprint{AIP/123-QED}
\title{Grid instability growth rates for explicit, electrostatic momentum- and energy-conserving particle-in-cell algorithms}
\author{Luke C. Adams*}%
\email{luke.adams-1@colorado.edu}%
\author{Gregory R. Werner}%
\affiliation{University of Colorado Boulder}%
\author{John R. Cary}
\affiliation{University of Colorado Boulder}%
\affiliation{Silvaco Inc.}%

\date{\today}

\begin{abstract}
When the Debye length is not resolved in a simulation using the most common particle-in-cell (PIC) algorithm, the plasma will unphysically heat until the Debye length becomes resolved via a phenomenon known as grid heating.
This article presents detailed numerical measurements of grid heating for several explicit PIC algorithms including the first systematic (covering the Debye length resolution and drift-velocity parameter space) study of grid-heating growth rates for the most-common electrostatic momentum-conserving PIC algorithm.
Additionally, we derive and test a cubic-spline-based PIC algorithm that ensures that the interpolated electric field has a continuous first derivative, but find that a differentiable electric field has minimal impact on grid-heating stability.
Also considered are energy-conserving PIC algorithms with linear and quadratic interpolation functions.
In all cases, we find that unphysical heating can occur for some combinations of Debye under-resolution and plasma drift.
We demonstrate analytically and numerically that grid heating cannot be eliminated by using a higher-order field solve, and give an analytical expression for the cold-beam stability limits of some energy-conserving algorithms.
\end{abstract}

\maketitle

\section{\label{sec:intro}Introduction}

The particle-in-cell (PIC) method has a long and successful history of simulating plasmas.\cite{hockney1966,birdsall2004}
In PIC methods, the particle distribution function is approximated by a sum of \emph{macroparticles}, each of which can occupy any location within the simulation domain.
In contrast, field quantities---including charge/current density and the electromagnetic fields---are defined on a discrete grid.
As a result, an interpolation function must be used to weight macroparticle charge/current to the grid and another interpolation function must be used to weight electromagnetic fields to the macroparticle locations.
With a uniform grid and identical interpolation functions, this method is momentum conserving, and so we will refer to it as MC-PIC.

From the outset, it was apparent that PIC methods suffered from a range of numerical instabilities,\cite{langdon1970a,langdon1973} but careful characterization of these instabilities has enabled practitioners to avoid the most problematic regions of instability.\cite{birdsall2004,barnes2021,powis2023}
For example, in relativistic electromagnetic simulations, the macroparticles can exceed the numerical speed of light (itself a consequence of the dispersion of the field advance), resulting in numerical Cherenkov radiation; modifications to the field update can suppress this instability.\cite{nuter2014}
Additionally, in electromagnetic PIC simulations, the finite timestep introduces inaccuracies in the interaction between the particles and fields resulting in violations of energy conservation.
This violation can be mitigated with more advanced time-integration schemes that ensure consistency between the field and particle advances.\cite{markidis2011,higuera2017,gonoskov2024}
Care must also be taken when depositing charge and current from particles to the grid to ensure that the electromagnetic fields sourced by the particles are consistent with the particle charge density.\cite{villasenor1992,esirkepov2001,umeda2003}

In this paper, we restrict our focus to electrostatic PIC simulations which calculate the fields self-consistently from the particle positions at each step.
However, electrostatic PIC simulations still suffer from instabilities.

In particular, we will focus on the grid-heating instability, which can rapidly and unphysically heat a plasma.\cite{langdon1973,birdsall2004}
This instability occurs when the grid spacing, $\Delta x$, is more than a few times the Debye length
\begin{equation}
    \lambda_D = \sqrt{\frac{\epsilon_0 k_B T}{n q^2}},
    \label{eq:debye_length}
\end{equation}
where $k_B$ is the Boltzmann constant, $T$ is the electron temperature, $n$ is the electron number density, and $q$ is the electron charge.
As the plasma heats, $\lambda_D / \Delta x = v_t / \omega_p \Delta x$ grows until it is order unity and the instability ceases.
Here, $v_t = \sqrt{k_B T / m}$ is the electron thermal velocity ($m$ is the electron mass) and $\omega_p$ is the electron plasma frequency.

The requirement to resolve the Debye length renders some physically relevant simulations infeasible.
For example, magnetron sputtering devices and Hall thrusters typically operate at very low plasma densities, which makes the kinetic simulation of full-device operation infeasible without approximations.\cite{taccogna2019,ryabinkin2021,reza2024,reza2024a}

We note that the grid-heating instability is distinct from the stochastic heating that occurs due to unphysically large fluctuations in the electric field resulting from the discretization of the distribution function.\cite{hockney1971,jubin2024}
These heating mechanisms can be distinguished by their behavior in the limit of an infinite number of macroparticles: the grid-heating growth rate will tend towards some limiting value while the stochastic heating rate will vanish in inverse proportion to the number of macroparticles per cell.\cite{acciarri2024,werner2025}

\begin{figure}
    \centering
    \includegraphics{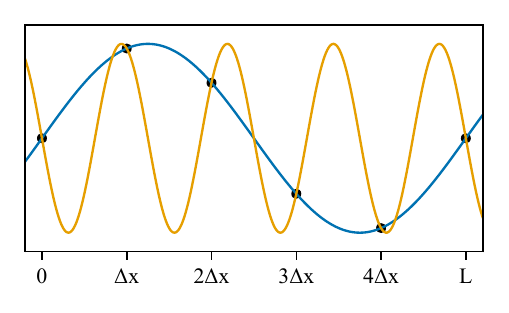}
    \caption{When the particle density varies with subgrid wavelength, the subgrid mode (orange) is unphysically aliased to the grid mode (blue).}
    \label{fig:aliasing}
\end{figure}

The source of the grid-heating instability can be understood by considering the Fourier modes of the interpolated particle density and the gridded quantities.
For any algorithm that relies on local interpolations between particles and the grid, subgrid modes (i.e.~modes with a wavelength shorter than can be resolved by the grid) in the particle density will alias onto grid modes during the charge-deposition interpolation (see \cref{fig:aliasing}).
The grid modes will then generate unphysical forces at the subgrid wavelengths due to aliasing in the field interpolation, and so these coupled modes can feed back on themselves and grow coherently.
When the Debye length is properly resolved, the subgrid modes are damped in a process analogous to Landau damping; underresolving the Debye length prevents this damping and leads to an instability and unphysical growth in the thermal energy.\cite{birdsall1980}

In an attempt to improve this, energy-conserving PIC  algorithms have been developed.\cite{birdsall2004,brackbill2016}
We consider a specific algorithm---first introduced by \textcite{lewis1970} and derived here in \cref{sec:ecpic}---that we will refer to as EC-PIC$i$ where $i + 1$ is the width in cells of the charge deposition interpolation.
It can be shown that this explicit algorithm conserves energy in the short timestep limit.\cite{lewis1970}

Energy-conserving algorithms simulating underresolved ($\lambda_D < \Delta x$) stationary plasmas are immune to the grid-heating instability because heating would violate energy conservation; however, if the plasma is both underresolved and has a bulk drift ($v_d \neq 0$), then energy-conserving algorithms may unphysically convert drift kinetic energy into thermal energy until the two energies are approximately equal.\cite{lewis1970,evstatiev2013}
This class of instability is often referred to as the cold-beam instability.
For the purposes of this paper, we will consider both cold-beam instability and the instability from underresolving the Debye length of a stationary plasma to be subclasses of the grid instability.

One result of the energy-conserving instability behavior is that, if a thermal plasma can be simulated in its rest frame where $v_d = 0$, the simulation will experience no heating.
Unfortunately, this is infeasible in cases where the plasma drift occurs within a larger system which fixes the simulation frame (e.g.~thin-film sputtering devices, Hall thrusters, and vacuum electronic devices),\cite{ryabinkin2021,han2017,taccogna2019,reza2024,reza2024a}.
Thus, the stability of these algorithms while simulating a drifting plasma remains of interest.
In these types of simulations, the plasma drift may be driven by an external energy source which may minimize the effect of the numerical heating.
However, such considerations are very simulation-dependent, and we will not address them in this paper.
Instead, we aim to provide general stability criteria that can indicate when simulation results should be subject to additional testing and scrutiny.

The exact threshold at which grid instability occurs depends on the algorithm used, and has been studied extensively in 1D.
For example, \textcite{birdsall1980} reported that, for a sufficiently large drift speed (somewhere around $v_d \gtrsim \Delta x \omega_p / 3)$, MC-PIC is unstable when $\lambda_D / \Delta x < 0.046$.
In the same paper,\cite{birdsall1980} Albritton and Nevins are credited with an analytical calculation showing that the threshold for stability for MC-PIC in the large-drift-velocity limit should occur around $\lambda_D / \Delta x \approx 0.05$.
Empirical estimates for the stability of stationary plasmas simulated using MC-PIC cluster around $\lambda_D / \Delta x = 1/\pi \approx 0.3$.\cite{birdsall1980,langdon1970,birdsall1975,birdsall2004}
We are not aware of any analytically-derived thresholds for the stability of stationary plasmas.

\textcite{barnes2021} have shown analytically that a hypothetical EC-PIC0 algorithm (which is not physically realizable for reasons that will be discussed later) would be stable for $v_d > \Delta x \omega_p / 2$ when $v_t = 0$.
They have additionally provided stability contours for 1D simulations using several energy-conserving algorithms as a function of $v_t$ (equivalently, Debye length resolution) and $v_d$.
Recently, energy-conserving methods have been benchmarked against the MC-PIC algorithm in the simulation of a capacitively-coupled discharge in both one and two dimensions;
coarse-resolution simulations with an energy-conserving algorithm showed excellent agreement with MC-PIC simulations that resolve the Debye length.\cite{turner2013,powis2023,sun2023}

When the Debye length is underresolved, the grid instability can be suppressed by implicit methods which allow for timesteps longer than an inverse plasma frequency.\cite{brackbill1982,barnes2021,li2023}
This stabilization occurs because the dispersion relation is modified to reduce the plasma frequency, which causes a corresponding reduction in the grid-instability growth rate.\cite{brackbill1982}
However, each timestep in an implicit method is very computationally expensive, although sometimes this is balanced by the reduced number of timesteps required.
It is possible to reduce the cost of implicit methods with certain approximations to the nonlinear system, but for grid-heating to be efficiently suppressed using these so-called direct implicit methods, the timestep must increase in proportion to the cell size.\cite{sun2023}
Thus, direct implicit methods cannot be used to suppress grid-heating in situations where temporal scale of interest is equal to or faster than the plasma frequency (e.g. long wavelength Langmuir waves).
Fully-nonlinear implicit algorithms do not suffer from this restriction, but they incur even worse performance penalties relative to explicit algorithms.\cite{chen2011}
We therefore consider techniques for reduction of grid-heating in cheaper and more easily implemented explicit algorithms.

In this article, we systematically measure the strength of the grid instability in several different algorithms---the standard momentum-conserving PIC algorithm (MC-PIC), the original energy-conserving algorithm proposed by \textcite{lewis1970} (EC-PIC1), a modified Lewis energy-conserving algorithm that uses wider basis functions for charge deposition and field interpolation (EC-PIC2), and a novel energy-conserving cubic-interpolating-spline based method which ensures that both the potential and its first derivative are continuous (CS-PIC).
We compare the growth rate as a function of drift and thermal velocity for each algorithm.

The remainder of the paper is organized as follows.
In \cref{sec:mcpic}, we introduce the PIC simulation cycle, and briefly discuss the MC-PIC algorithm.
The reader who is already familiar with this commonly-used PIC algorithm can safely skip this section.
Then in \cref{sec:energy_conserving}, we show how the Low Lagrangian\cite{low1958,evstatiev2013} can be used to derive energy-conserving algorithms.
\Cref{sec:ecpic} and \Cref{sec:cspic} derive EC-PIC and CS-PIC, respectively, with a focus on the commonality between the algorithms.
\Cref{sec:results} presents results of our systematic numerical study of these algorithms, while \cref{sec:cold_beam} presents new analytical results about the stability of the EC-PIC algorithms.
Finally, we offer some concluding remarks in \cref{sec:conclude}.

\section{The PIC simulation cycle} \label{sec:mcpic}
A collisionless plasma with one spatial dimension and one velocity dimension is described by the Vlasov equation
\begin{equation} \label{eq:vlasov}
    \pdv{f}{t} + v \pdv{f}{x} + \frac{q}{m} E \pdv{f}{v} = 0,
\end{equation}
where $f(x, v)$ is the particle distribution function for a species with mass $m$ and charge $q$ and where the electric field, $E$, is self-consistently calculated as
\begin{equation} \label{eq:gauss}
    \dv{E}{x} = \frac{1}{\epsilon_0} \int \dd{v} q f(x, v),
\end{equation}
and $\epsilon_0$ is the permittivity of free space.
For simplicity, we consider one  mobile species; a more complete description would have one distribution function per species.

In order to do simulations, the Vlasov-Poisson system defined in \cref{eq:vlasov,eq:poisson} must be discretized.
The MC-PIC method makes a series of ad hoc approximations which we now review.

In PIC methods, the particle density function is approximated by the sum
\begin{equation}
    f({x}, {v}) \approx \sum_{m=1}^M w_m \delta({x} - {x}_m) \, \delta({v} - {v}_m)
    \label{eq:dist_func_disc}
\end{equation}
where the sum ranges over the $M$ macroparticles in the simulation each with weight $w_m$ and phase-space coordinate $({x}_m, {v}_m)$, and where $\delta$ is the Dirac delta function.

We consider a system with length $L$, and with periodic boundary conditions.
The field quantities are defined on a grid, which we choose to have $N$ equally spaced nodes.
The locations of the nodes are given by $X_n = n \Delta x$ where $\Delta x = L / N$.

\subsection{Charge deposition} 
For the MC-PIC algorithm, the charge density at each grid node is computed as
\begin{align} \label{eq:charge_dep}
\rho_n
    &= q \int \dd{x} \int \dd{v} f(x, v) S_n(x) \\
    &= q \sum_m w_m S_n(x_m),
\end{align}
where $S_n(x)$ is the \emph{shape function} used to interpolate from a particle at position $x$ to grid node $n$.
The shape function can either be associated with the grid nodes, as we have done here, or it can be equivalently implemented by giving the particles a finite width by modifying the delta function in \cref{eq:dist_func_disc}.

However, the derivation of more complicated PIC algorithms (e.g. variable-sized grids) requires that a shape function be associated with the grid to ensure charge conservation.
In fact, it is possible for both the particles and grid nodes to be given a width, although this only serves to complicate the math.\cite{langdon1970}
In the forthcoming derivations, we will find it convenient to assume that the particles have infinitesimal width, and so we associate the shape functions with the grid nodes.

For a macroparticle's total charge to be conserved on the grid, we must choose $S_n$ such that
\begin{equation}
    \sum_{n} S_n(x) = 1
\end{equation}
for all $x$ in the simulation domain.
Additionally, the shape function $S$ should be \emph{local} (spatial extent of only a few grid cells) for the algorithm to scale efficiently as the number of grid cells increases.

\begin{figure}
    \centering
    \includegraphics{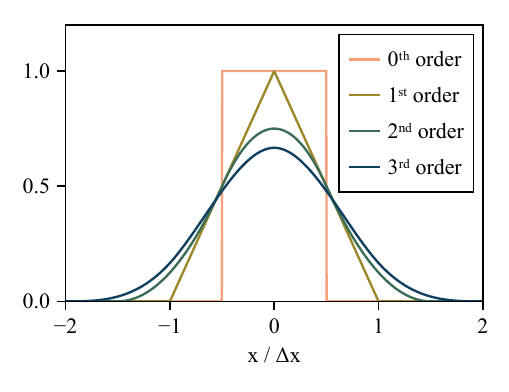}
    \caption{The first few B-splines. Higher-order B-splines are smoother (in the sense of approaching a Gaussian) but extend further.}
    \label{fig:bsplines}
\end{figure}

By far the most common choices for shape functions are B-splines, the first few of which are shown in Figure \ref{fig:bsplines}.
The use of zeroth-order B-splines for the interpolation is referred to as nearest grid point (NGP) interpolation, while the use of first-order B-splines is often called cloud-in-cell (CIC).\cite{birdsall2004}
In this paper, MC-PIC refers to first-order interpolation.

\subsection{Field solve}
Once the charge density on the grid has been found, the electric field must be computed.
For the MC-PIC algorithm, this is done in two steps: first find the potential due to the charge by solving a discretized Poisson's equation, and then compute the electric field at each node as a finite-difference derivative of the potential.
We now consider each of these steps in detail.

In one dimension, the continuous Poisson equation is given by
\begin{equation} \label{eq:poisson}
    -\dv[2]{\phi}{x} = \frac{\rho}{\epsilon_0}.
\end{equation}
This equation can be discretized using the three-point differencing formula
\begin{equation} \label{eq:three-point}
    \frac{-\phi_{n - 1} + 2 \phi_n - \phi_{n+1}}{\Delta x^2} = \frac{\rho_n}{\epsilon_0}
\end{equation}
where the indices of $\phi$ are taken to be periodic so that $\phi_0 = \phi_N$.
This is a standard problem in linear algebra, and many exact and approximate techniques exist to calculate the solution.\cite{golub2013,numRecipes3rd}

The potential is defined on the grid nodes, and in a typical PIC code, the electric field is also interpolated from the grid nodes to the particle locations.
Thus, a simple finite difference to compute the electric field is a centered difference given by
\begin{equation}
    E_n = - \frac{\phi_{n+1} - \phi_{n-1}}{2 \Delta x}.
\end{equation}
This centered difference is often interpreted as a two-step process. First, the electric field at the grid edges is computed as
\begin{equation} \label{eq:edge_E_field}
    E_{n + 1/2} = - \frac{\phi_{n+1} - \phi_n}{\Delta x},
\end{equation}
and then the edge electric fields are averaged to give a nodal electric field
\begin{equation}
    E_n = \frac{E_{n + 1/2} + E_{n - 1/2}}{2}.
\end{equation}
For a uniform grid, these two approaches are identical.

\subsection{Field interpolation}
In MC-PIC, the force acting on each macroparticle is computed using the same shape function that was used for charge deposition.
Thus,
\begin{equation} \label{eq:force}
    F(x_m) = q w_m \sum_n E_n S_n(x_m).
\end{equation}
This choice ensures momentum conservation provided that the chosen field solve is symmetric.

Again, $S_n$ is local, and so the evaluation of this sum for all macroparticles can be done in $O(M)$ time (i.e.~the scaling does not depend on $N$).

\subsection{Particle push} \label{sec:particle_push}
Finally, the positions and velocities of the macroparticles must be updated.
The simplest explicit, electrostatic time integrator with acceptable accuracy is the leapfrog integrator.
In a leapfrog integrator, the positions of the macroparticles are stored at whole timesteps $i \Delta t$, while the velocities of the macroparticles are stored at the half timesteps, $(i + 1/2) \Delta t $.
The velocities of the macroparticles are first updated as
\begin{equation}
    v_m^{i + 1/2} = v_m^{i - 1/2} + \Delta t \frac{F(x_m^{i})}{w_m m}
\end{equation}
Then the positions of the macroparticles are updated according to
\begin{equation}
    x_m^{i + 1} = x_m^{i} + \Delta t v_m^{i + 1/2}.
\end{equation}
We note that only the most recent position and velocity of each particle must be stored.

Although MC-PIC generally provides a good approximation to \cref{eq:vlasov}, it is not guaranteed to preserve all of the invariants present in the original equations.
It can be shown that the preceding algorithm exactly conserves momentum.\cite{birdsall2004}\textsuperscript{,}\footnote{The important pieces required for this conservation are a symmetric field solve, and identical interpolations for the charge deposition and field interpolation.}
However, MC-PIC does not strictly conserve energy, although it becomes increasingly good in the limit that $N \to \infty$, $\lambda_D / \Delta x \to \infty$, and $\omega_p \Delta t \to 0$.
In the next sections, we will derive several algorithms that \emph{do} exactly conserve energy (for an infinitesimal timestep), although we will find that we must sacrifice momentum conservation in order to maintain good performance of the algorithms.

\section{Derivations of energy-conserving algorithms} \label{sec:energy_conserving}
The starting point for all of our derivations of energy-conserving PIC algorithms will be the Low Lagrangian,\cite{low1958,evstatiev2013} which is given by
\begin{multline} \label{eq:low_lagrangian}
	\mathcal{L} = \iint \dd{x_0} \dd{v_0} f_0(x_0, v_0) \left[
        \frac{m}{2} \left(\pdv{x(t; x_0, v_0)}{t}\right)^2 \right. \\
        \left. \vphantom{\left(\frac12\right)^2} - q \phi(x(t; x_0, v_0), t) \right]
    + \frac{\epsilon_0}{2} \int \dd{\xi} \left( \pdv{\phi(\xi, t)}{\xi} \right)^2.
\end{multline}
Given some initial phase-space density $f_0(x_0, v_0)$, the function $x(t; x_0, v_0)$ gives the trajectory of the initial phase-space coordinate $(x_0, v_0)$.
The first term of the Lagrangian describes the particle kinetic energy, the second term encodes the particle-field coupling and the last term is the electric field energy.
We use the variable $\xi$ to distinguish the integration variable from the phase-space coordinate $x$.

The degrees of freedom of the Lagrangian are the fields $x(x_0, v_0)$ and $\phi(x)$.
The Euler-Lagrange equations for $x$ can be calculated as\cite{evstatiev2013}
\begin{equation}
    \fdv{\mathcal{L}}{x} = \pdv{t} \fdv{\mathcal{L}}{(\partial_t x)} + \pdv{x_0} \fdv{\mathcal{L}}{(\partial_{x_0} x)} + \pdv{v_0} \fdv{\mathcal{L}}{(\partial_{v_0} x)}
\end{equation}
which yields
\begin{equation} \label{eq:low_newton}
    - q \pdv{\phi}{x} = m \ddot{x} + 0 + 0.
\end{equation}
The equations of motion for $\phi$ are calculated as
\begin{equation}
    \fdv{\mathcal{L}}{\phi} = \pdv{t} \fdv{\mathcal{L}}{(\partial_t \phi)} + \pdv{x_0} \fdv{\mathcal{L}}{(\partial_{x_0} \phi)} + \pdv{v_0} \fdv{\mathcal{L}}{(\partial_{v_0} \phi)}
\end{equation}
yielding
\begin{equation} \label{eq:low_poisson}
    -q \int \dd{v} f(x, v) = 0 + \epsilon_0 \pdv[2]{\phi}{x} + 0.
\end{equation}
\Cref{eq:low_newton,eq:low_poisson} exactly reproduce \cref{eq:vlasov,eq:gauss}.
Thus, this Lagrangian preserves the symmetries responsible for energy and momentum conservation.
That is, the Lagrangian is invariant to translations in time or space.
It is the discretization of the Lagrangian that breaks the symmetries, leading to violations of conservation laws.

To derive a PIC-like algorithm, we substitute the macroparticle approximation into the distribution function given in \cref{eq:dist_func_disc}.
The resulting Lagrangian is
\begin{equation} \label{eq:particle_lagrange}
	\mathcal{L} = \sum_m w_m \left[ \frac{m}{2} \dot{x}_m^2 - q \phi(x_m) \right]
    + \frac{\epsilon_0}{2} \int \dd{\xi} \left( \pdv{\phi}{\xi} \right)^2,
\end{equation}
where the field degree of freedom $x$ has been replaced by the $M$ degrees of freedom $\{x_m\}$.
This new Lagrangian retains continuous space and time symmetries, and so it will conserve both energy and momentum.
However, we still must choose a discretization for the potential $\phi$.

For the resulting algorithm to have PIC-like scaling, the total cost of the algorithm cannot have a term that scales as $O(M N)$.
This requires each macroparticle to interact with a limited number of grid nodes, even as the number of grid cells is increased.
In the next two sections, we consider two different approaches to discretizing the potential that respect this requirement.

\section{Energy-conserving PIC (EC-PIC)} \label{sec:ecpic}
Following \textcite{lewis1970}, we discretize the potential as a sum of local basis functions
\begin{equation}
    \phi(x) \approx \sum_{n=1}^N \phi_n \, S \! \left(\frac{x - X_n}{X_n - X_{n-1}}\right)
    \equiv \sum_{n=1}^N \phi_n \, S_n(x),
    \label{eq:pot_bspline}
\end{equation}
where $S$ is a local basis function (shape function), and the coordinates $X_n$ are the locations of grid nodes.
Thus, the $S_n(x)$ functions form a set of basis functions that span the domain, each localized around $X_n$.

The resulting Lagrangian is
\begin{equation}
\begin{aligned}
    L_\text{EC} &= \sum_m \frac{w_m m}{2} \dot{x}_m^2 \\
    &- \sum_m w_m q \sum_n \phi_n \, S_n(x_m) \\
    &+ \frac{\epsilon_0}{2} \sum_n \sum_\ell \phi_n \left[ \int \dd{x} S_n'(x) S_\ell'(x) \right] \phi_\ell,
    \label{eq:bspline_lagrangian}
\end{aligned}
\end{equation}
where $n$ and $\ell$ independently vary over the grid nodes, and the prime indicates a derivative with respect to the argument.
The equations of motion for $x_m$ and $\phi_n$ are
\begin{align}
    w_m m \ddot{x}_m &= - w_m q \sum_n \phi_n S_n'(x_m) \label{eq:ec_particle_eom} \\
    - [\nabla^2]_{n\ell} \phi_n &= \frac{1}{\epsilon_0} \sum_m w_m q S_n(x_m) \equiv \frac{\rho_n}{\epsilon_0}, \label{eq:finite_field_solve}
\end{align}
where $[\nabla^2]_{n\ell}$ is a finite-difference approximation of the Laplacian operator which can be computed as
\begin{equation} \label{eq:lagrangian_field_solve}
    [\nabla^2]_{n\ell} = \int \dd{x} S_n'(x) S_\ell'(x).
\end{equation}
Recall that the shape function $S$ has been chosen to be local in \cref{eq:bspline_lagrangian}, and thus the elements of the operator will be nonzero only near the diagonal (i.e., $n \approx \ell$).
The equations of motion \labelcref{eq:ec_particle_eom,eq:finite_field_solve} represent a fully spatially discretized approximation to the Vlasov-Poisson system.

We now compare this energy-conserving algorithm with the MC-PIC algorithm presented in \cref{sec:mcpic}.
For simplicity, we specialize to the case of a uniform grid, and choose $S$ to be an $i$th order B-spline.

\subsection{Charge deposition} By defining $\rho_n$ as in \cref{eq:finite_field_solve}, the charge deposition is the same as MC-PIC (\cref{eq:charge_dep}).

\subsection{Field solve} The discretized Poisson's equation is given by \cref{eq:finite_field_solve}.
The structure of the equation is similar to the MC-PIC field solve; however, the details of the field solve now depend on the inner products of shape functions.
We note that the derivative of $S$ is required to compute the discretized Laplacian, and thus the zeroth-order B-spline is unsuitable for energy-conserving PIC algorithms.%

If $S$ is chosen to be the first-order B-spline, then it can be shown that the discretized Laplacian becomes tridiagonal (plus the usual periodic off-diagonal elements), with diagonal elements equal to $-2 / \Delta x^2$, and off-diagonal elements equal to $1 / \Delta x^2$.
Thus, for EC-PIC1 (i.e., charge deposition with first-order B-splines), the field solve is identical to that of the MC-PIC scheme (\cref{eq:three-point}).

Alternatively, wider shape functions can be used to reduce particle noise, but this choice also comes with some trade-offs.
First, computing the charge deposition becomes more expensive as each particle must deposit charge to additional grid nodes.
Additionally, the introduction of a wider stencil can slow down the speed of simulations, both by increasing the number of nonzero elements in the field linear solve, and by requiring the communication of additional guard cells when parallelizing the algorithm.
Further, the application of boundary conditions becomes more challenging as more grid nodes interact with the edges of the grid.
In this paper, we avoid this complication by utilizing periodic boundary conditions.

For EC-PIC2 (second-order B-spline charge deposition), the difference equations to be solved are
\begin{equation} \label{eq:poisson-lagrange}
    \frac{\phi_{n-2} + 2 \phi_{n-1} - 6 \phi_n + 2 \phi_{n+1} + \phi_{n+2}}{6 \Delta x^2} = \frac{\rho_n}{\epsilon_0}.
\end{equation}
and so the discretized Laplacian matrix has five nonzero diagonals, with rows consisting of $[1/6, 1/3, -1, 1/3, 1/6] / \Delta x^2$.
We note that the $\phi_n$ in \cref{eq:poisson-lagrange} are the amplitudes of the potential elements defined in \cref{eq:pot_bspline} and not the value of the potential at each grid node.
We refer to this algorithm as EC-PIC2-Lagrange.

We additionally consider the EC-PIC2 algorithm with two modified field solves.
The first solve, which we refer to as EC-PIC2-Standard, uses the standard three-point field solve employed by MC-PIC (\cref{eq:three-point}) and EC-PIC1, with second-order error in $k \Delta x$. 
This scheme may be derived in the Lagrangian formulation by using a zeroth-order B-spline basis for the potential in \cref{eq:pot_bspline} paired with a modified version of the distribution function discretization, \cref{eq:dist_func_disc}, that gives each particle a zeroth-order B-spline spatial extent.

We also consider a fourth-order-accurate, five-point field solve with rows consisting of $[1 / 12, -4/3, 5/2, -4/3, 1/12]/\Delta x^2$.\cite{abramowitz1965}
We refer to this algorithm as EC-PIC2-Fourth.
We are not aware of a derivation of the EC-PIC2-Fourth from \cref{eq:low_lagrangian}; however, the scheme can be derived using a variation of the Hamiltonian procedure in \textcite{birdsall2004}, Ch. 10.

Although the field solve in the EC-PIC2-Lagrange algorithm has been specifically derived to be consistent with the 2nd-order B-spline potential representation, the field solve itself is actually more inaccurate than the other two field solves for $k \Delta x \sim \pi$.
However, it has been shown that, considering the overall algorithm, using this field solve will result in every particle density mode oscillating at exactly the cold plasma frequency.\cite{birdsall2004}

\subsection{Field interpolation} The interpolation of the field to the macroparticles is carried out in the same way as in MC-PIC (\cref{eq:force}), except that the shape function $S_n$ is replaced with the derivative $S_n'$ and the edge electric field values are used in place of the nodal electric fields.
If the shape function $S$ is an $i$th order B-spline, then this is equivalent to interpolating the edge electric fields defined in \cref{eq:edge_E_field} with a B-spline shape interpolation of order $(i - 1)$.\cite{birdsall2004}

\subsection{Particle push} The equations of motion, \labelcref{eq:ec_particle_eom,eq:finite_field_solve}, have not yet been discretized in time, and thus many different energy-conserving time integration schemes may be employed to advance the particle positions and momenta.
In this paper, we use the same explicit leapfrog integrator described in \cref{sec:particle_push}.

Prior to the discretization of the potential, the Lagrangian was translationally invariant, and thus by Noether's theorem it had a conserved quantity---momentum.
Discretization breaks this symmetry, and so the momentum is no longer guaranteed to be conserved.
One might hope that momentum remains a conserved quantity despite the lack of symmetry; however, it has been shown explicitly that EC-PIC lacks momentum conservation.\cite{birdsall2004,evstatiev2013}

In contrast, since it is not derived from a Lagrangian, MC-PIC does not have this symmetry restriction.
It can be shown that MC-PIC conserves momentum even though it lacks translational invariance.\cite{birdsall2004}
This conservation comes as a consequence of Newton's 2nd law, and not from a symmetry of the equations of motion.

\section{Cubic-spline PIC (CS-PIC)} \label{sec:cspic}
We are motivated by the Particle-in-Fourier (PIF) method,\cite{evstatiev2013,mitchell2019} which has been shown to conserve both energy and momentum exactly in the short timestep limit.\cite{evstatiev2013}
In PIF, $\phi$ is represented as a truncated Fourier series, with every particle depositing charge to every Fourier mode.
Such an algorithm does not have PIC-like scaling, and so the simulation of large-scale problems is infeasible.

PIC-like scaling can be recovered by using approximate unequally-spaced fast Fourier transforms;\cite{mitchell2019} however, this approach does not scale to more complicated boundary conditions, and it requires sophisticated numerical algorithms that are not widely available in numerical computing packages.\cite{beylkin1995}
Additionally, FFTs present a challenge for parallel processing, and do not generalize well to higher dimensions.

Since the Fourier mode discretization exactly conserves momentum, we hypothesize that basis functions that approximate Fourier modes will provide a superior approximation to momentum conservation.
To this end, we consider an alternative discretization of the potential which uses cubic interpolating splines to approximate the Fourier modes.
We will see that this choice of approximation allows for PIC-like scaling of the deposition and field interpolations.

A cubic interpolating spline is a piecewise-cubic function that is constructed to pass through a set of values defined on grid points.
Unlike a B-spline interpolation, which is explicitly specified by the grid point values, a cubic interpolating spline generally requires solving a tridiagonal linear system for the values of the second derivative of the interpolation at each grid point.
The resulting interpolation is continuous and has continuous first and second derivatives.\cite{numRecipes3rd}
To fully specify the linear solve, the second derivatives at each end of the interpolation domain must be specified.
For the purposes of this paper, which only considers periodic boundary conditions, the system can be closed by requiring that the derivatives at each end point match.

If the electric potential is represented using a cubic interpolating spline, then the interpolated electric field will have a continuous first derivative.
In the EC-PIC method, this only becomes true for third-order and higher-order B-splines, with the associated computational cost of interpolating to and from more cells for each particle.
Thus, unlike the EC-PIC methods considered in this paper, CS-PIC ensures that small translations in the location of the grid will result in only small changes to the electric field experienced by macroparticles.

Before we describe the derivation of the method, we briefly review cubic interpolating splines.
Consider a periodic system of length $L$.
We impose the same $N$ node grid with nodes at $X_n = nL/N$.
For some field $g$ with grid values $g(X_n)$, the cubic interpolation, $g_C$, is given by
\begin{equation} \label{eq:cs_interp}
	g_C(x) = \sum_{n=0}^{N} \left( g(X_n) w_1(\xi_n) +
	g''(X_n) \Delta x^2 \, w_3(\xi_n) \right),
\end{equation}
where $\xi_n = (X_n - x)/ \Delta x$ is the normalized distance from grid point $n$ and weight functions are defined as
\begin{equation} \label{eq:weight1}
	w_1(\xi) =
    \begin{cases}
        1 - |\xi| & |\xi| < 1\\
        0 & |\xi| \geq 1\\
    \end{cases}
\end{equation}
and
\begin{equation} \label{eq:weight3}
	w_3(\xi) =
    \begin{cases}
       -\frac13 |\xi| + \frac12 |\xi|^2 - \frac16 |\xi|^3 & |\xi| < 1\\
        \phantom{-}0 & |\xi| \geq 1\\
    \end{cases}
\end{equation}
The weight functions are only nonzero on $|\xi| < 1$, and so the evaluation of $f_C(x)$ can be done in $O(1)$ time for arbitrarily large $N$.

We differentiate \cref{eq:cs_interp} with respect to $x$---evaluated at $X_n$---and require continuity to obtain\cite{numRecipes3rd}
\begin{equation} \label{eq:cs_deriv_solve}
\begin{split}
    \frac{g''(X_{n - 1})}{6} + & \frac{2g''(X_n)}{3} + \frac{g''(X_{n + 1})}{6} = \\
    & \frac{g(X_{n - 1}) - 2 g(X_n) + g(X_{n + 1})}{\Delta x^2}.
\end{split}
\end{equation}
Using this relation, the values of the second derivatives $g''(X_n)$ can be found by solving the tridiagonal linear problem \cref{eq:cs_deriv_solve} with a source term that only depends on the function values $g(X_n)$.

We consider a cubic spline approximation to a sinusoidal mode with nodal values $e_k(X_n) = \exp(ikX_n)$, where $k$ is the wavenumber of a grid mode (i.e.~$k = 2\pi n/L$ with $-N/2 + 1 < n \leq N / 2$).
We make the ansatz that the values of the second derivative will also vary sinusoidally with the same $k$: that is $e''(X_n) = \hat{C}_k \exp(ikX_n)$.
It can then be shown that
\begin{equation} \label{eq:c_k}
    \hat{C}_k = - k^2 \sinc^2\left(\frac{k\Delta x}{2}\right)\frac{3}{2 +
	\cos(k\Delta x)},
\end{equation}
with $\sinc(x) = \sin(x) / x$.
Thus, for the special case of sinusoidal modes, the second-derivative values can be calculated without requiring a tridiagonal solve.
\Cref{eq:c_k} reduces to the expected continuum result in the $k\Delta x \to 0$ limit.

Because both $e_k(X_n)$ and $e''(X_n)$ have an $\exp(ikX_n)$ dependence, it can be shown that
\begin{equation}
    \langle e_k | e_{\ell} \rangle = \int_0^L \dd{x} e_k^*(x) e_{\ell}(x) \propto \delta_{k,\ell}
\end{equation}
where $\delta_{k, \ell}$ is a Kronecker delta function.
That is, the $e_k$'s are orthogonal just like the Fourier modes that they approximate.
The details of this calculation are shown in the Appendix.

Following the derivation of PIF,\cite{evstatiev2013,mitchell2019} we discretize $\phi$ as
\begin{equation} \label{eq:pics_phi}
    \phi(x) \approx \sum_{k=-N/2 + 1}^{N/2} \hat{\phi}_k e_k(x),
\end{equation}
where $e_k$ is an cubic interpolating cubic spline with $e_k(X_n) = \exp(ikX_n)$.
The Lagrangian becomes
\begin{equation}
\begin{aligned}
    L_\text{CS} &= \sum_m \frac{w_m m}{2} \dot{x}_m^2 \\
    &- \sum_m w_m q \sum_k \hat\phi_k \, e_k(x_m) \\
    &+ \frac{\epsilon_0}{2} \sum_k \sum_\ell \hat\phi_k^* \left[ \int \dd{x} e_k'^{*}(x) e_\ell'(x) \right] \hat\phi_\ell.
\end{aligned}
\end{equation}

The Appendix shows that the inner products between derivatives of the basis functions appearing in the previous equation are also orthogonal, with values
\begin{equation} \label{eq:cspic_field_solve_exact}
    \langle e'_k | e'_\ell \rangle
    = \frac{L}{\Delta x^2} \frac{48 - 9c - 36 c^2 - 3 c^3}{5 (2 + c)^2} \delta_{k,\ell}
    \equiv k^2 L \hat{D}_k \delta_{k,\ell},
\end{equation}
where $c = \cos(k \Delta x)$.
In the limit of small $k \Delta x$, the correction $\hat{D}_k$ goes to one, and this reduces to the sinusoidal result: $\langle e'_k | e'_\ell \rangle \approx k^2 L \delta_{k,\ell}$.
The Euler-Lagrange equations for $x_m$ and $\hat\phi_k$ are therefore
\begin{align}
    w_m m \ddot{x}_m &= -q_m\sum_k \hat{\phi}_k e'_k(x) \label{eq:cspic_push}, \\
    \hat\phi_k &= \frac{q}{\epsilon_0 k^2 L \hat{D}_k} \sum_m w_m e_k(x_m) \equiv \frac{\hat\rho_k}{\epsilon_0 k^2 \hat{D}_k} \label{eq:cspic_solve}.
\end{align}
A naive evaluation of \cref{eq:cspic_solve} would require calculating $e_k(x_m)$ for every $k$ and every $m$, and thus would scale as $O(M N)$.
However, we can exploit the properties of the cubic interpolating splines to evaluate the sums with $O(M)$ work.

\subsection{Charge deposition}
The charge density, $\hat\rho_k$, is defined in \cref{eq:cspic_solve} and can be efficiently computed as
\begin{align}
	\hat\rho_k
    &= \frac{q}{L} \sum_m w_m e_k(x_m) \\
    &= \frac{q}{L} \sum_m \sum_n w_m \left( e^{ikX_n}w_1(\xi_{n,m}) + \hat{C}_k
		e^{ikX_n}\Delta x^2 w_3(\xi_{n,m}) \right) \\
	&= \frac1N \sum_n \left( e^{ikX_n} \rho_n + \hat{C}_k e^{ikX_n} \rho''_n \right)
\end{align}
where we have defined \emph{two} charge density fields
\begin{align}
    \rho_n   &= q \sum_m \frac{w_m}{\Delta x} \, w_1(\xi_{n,m}) \\
    \rho''_n &= q \sum_m w_m \Delta x \, w_3(\xi_{n,m}),
\end{align}
and where $\xi_{n,m} = (X_n - x_m) / \Delta x$.
The computations of $\rho_n$ and $\rho''_n$ have PIC scaling and $\hat\rho_k$ can be efficiently computed using FFTs.
Crucially, the macroparticles are only ever involved in the local deposition to the two charge density fields, and the field solve has the same scaling in number of cells as MC-PIC and EC-PIC.

\subsection{Field solve}
The charge density weights $\hat{\rho}_k$ have already been computed in the previous step.
This means that potential weights $\hat{\phi}_k$ can be directly computed in \cref{eq:cspic_solve}.


\subsection{Field interpolation}
The field interpolation can be rewritten as
\begin{align}
	w_m m \ddot{x}_m
	&= \frac{q_m}{\Delta x}\sum_k \sum_{n=0}^{N} \hat{\phi}_k \left(
		e^{ikX_n} w'_1(\xi_{n,m}) + \hat{C}_k e^{ikX_n} \Delta x^2 \, w'_3(\xi_{n,m})
		\right) \\
	&= \frac{q_m}{\Delta x} \sum_{n=0}^{N} \left(
		\phi_n w'_1(\xi_{n,m}) + \phi''_n \Delta x^2 \, w'_3(\xi_{n,m})
		\right),
\end{align}
where the \emph{two} potential fields are calculated as
\begin{align}
	\phi_n   &= \sum_{k} \hat{\phi}_k e^{ikX_n}, \\
	\phi''_n &= \sum_{k} \hat{C}_k \hat{\phi}_k e^{ikX_n}.
\end{align}
Once again, the field interpolation is local, and so the CS-PIC algorithm can be implemented with PIC scaling.
Additionally, the use of FFTs allows for a direct computation of $\phi''_n$ field without requiring an additional linear solve, as is usually required for cubic interpolation splines.
Moreover, since the potential, as a cubic-interpolating spline, has a continuous second derivative, the interpolated electric field is guaranteed to have a continuous first derivative.

\subsection{Particle push}
The particle push is unaffected by the choice of discretization for the potential.
We once again use the leapfrog integrator described in \cref{sec:particle_push} for simplicity and consistency.

We note in passing that the potential discretization in \cref{eq:pics_phi} has fourth-order error, and so we expect that CS-PIC will have fourth-order error in $k\Delta x$.

\section{Simulation results} \label{sec:results}
The grid instability occurs when the Debye length of a plasma is underresolved.
We parameterize this using a normalized thermal velocity, $\bar{v}_t \equiv v_t / \omega_p \Delta x = \lambda_D / \Delta x$.
The grid instability growth rate additionally depends on the normalized drift velocity of the plasma, $\bar{v}_d \equiv v_d / \omega_p \Delta x$.

In general, the growth rate may also depend on the number of particles per cell, and on the timestep used.
However, analytical descriptions of grid instability typically assume continuous particle distributions and time evolution.
Thus, in our numerical investigations of grid heating, we use $2^{14} = 16,384$ particles per cell to ensure agreement with analytical descriptions.\cite{birdsall1980,birdsall2004}
Additionally, the use of a large number of particles per cell reduces the noise-driven heating---sometimes called stochastic heating---of the plasma, enabling measurements of small instability growth rates.\cite{werner2025}

The energy-conserving algorithms presented in this paper are only approximately energy-conserving for a finite timestep.
Thus, to ensure that the results would be applicable to real world simulations, we chose to simulate with $\omega_p \Delta t = 0.5$.
We have verified that the measured growth rates do not change substantially if a shorter timestep is used.

To compare the grid-instability behavior of the previously described methods (MC-PIC, EC-PIC1, EC-PIC2, CS-PIC), we conducted a series of periodic-boundary simulations with one spatial dimension and one velocity dimension.
The simulations were run using \texttt{ParticleInCell.jl}\footnote{\url{https://github.com/JuliaPlasma/ParticleInCell.jl}}, a PIC code implemented in the Julia\cite{julia2017} language and designed for testing novel PIC algorithms.
All plots are produced using the \texttt{Makie.jl} package.\cite{danisch2021}

Each simulation consisted of 64 cells initialized with a drifting thermal electron plasma neutralized by a stationary background density representing the ions.
The simulations used a quiet start to eliminate particle noise, following the procedure outlined in \textcite{birdsall2004}, Chapter 16, and briefly described here.
In each cell, particle velocities were assigned such that each particle represented an equal area under the Maxwellian distribution.
The particle positions within each cell were generated using a bit reversed sequence, which produces a deterministic scramble of the previously ordered velocities.
Thus, for a simulation with $M_\text{ppc}$ particles per cell, the phase space will consist of $M_\text{ppc}$ beams, each with exactly one particle per cell.

With this initial particle distribution, the initial electric field is zero and all of the energy is in the particle thermal and drift-kinetic energies.
Additionally, the electric field should remain zero even after the particles begin to evolve.\cite{gitomer1976}
However, eventually floating-point truncation noise will cause the beams to interact, and the electric field will become nonzero.
To speed this process and ensure that the most unstable mode will be seeded with a perturbation, we perturb the particle velocities at all wavenumbers with an amplitude of $10^{-8} \omega_p \Delta x$ and with a random phase. 

Since the initial condition consists of many beams, care must be taken to ensure that a multi-beam instability does not obscure the grid-instability growth.\cite{gitomer1976}
To guard against this possibility, we use many particles per cell, and monitor for an energy exchange from the particles to the fields, which indicates a multi-beam instability.

\begin{figure}
    \includegraphics{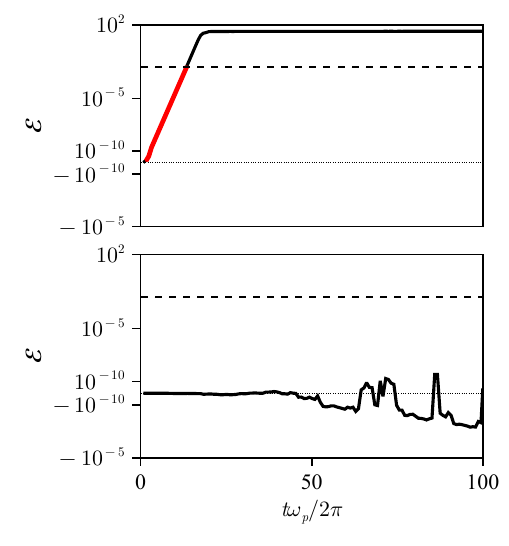} \\
    \caption{%
        Example growth rate measurements for an unstable (EC-PIC1, $\bar{v}_t = 0.01$, $\bar{v}_d = 0.05$, top panel) and a stable (EC-PIC1, $\bar{v}_t = 0.1$, $\bar{v}_d = 0.05$, bottom panel) simulation.
        The normalized change in thermal energy, $\mathcal{E}$, is shown as a function of time on a linear scale for $|\mathcal{E}| < 10^{-10}$ and a symmetric log scale otherwise.
        The portion of the time series up to and including the last negative value is discarded for the fitting procedure.
        Additionally, the portion of the time series that exceeds some cutoff---$1\%$ here---is discarded to isolate the region of true exponential growth in $\mathcal{E}$.
        A line is fit to this region of the time series (shown in red), and the grid heating growth rate is half the slope of this fit.
        If there is no remaining portion of the time series, as in the bottom panel, then the simulation is assumed to be stable.
    }
    \label{fig:growth_rate_plot}
\end{figure}

At each timestep, the total electron thermal energy was calculated as
\begin{equation}
    E_\text{th}(t) = \sum_m \frac12 w_m m (v_m(t) - \bar{v}(t))^2
    \label{eq:electron_thermal_energy}
\end{equation}
where $\bar{v}(t)$ is the average electron velocity of all electrons in the simulation at time $t$.
The normalized change in thermal energy at time $t$ is then given by $\mathcal{E}(t) \equiv (E_{\text{th}}(t) - E_\text{th}(0)) / E_\text{th}(0)$.
If $\mathcal{E}$ becomes negative with a magnitude above the noise floor during a simulation, this indicates that a multi-beam instability has occurred.
Also, at early times, $\mathcal{E}$ can become negative due to floating-point truncation noise.
We therefore discard the portion of the time series before and including the last negative value of $\mathcal{E}$.

For each simulation, a line was fit to $\ln(\mathcal{E})$ for the period before $\mathcal{E}$ exceeded $10^{-2}$.
The cutoff ensures that only the period of active exponential growth is fit.

Any fit with a correlation coefficient ($r^2$) of less than 0.9 was assumed to have zero slope.
The instability growth rate is half of the measured thermal energy growth rate.
\Cref{fig:growth_rate_plot} shows this fitting process for examples of stable and unstable simulations.

\begin{table*}
    \caption{Properties of tested PIC algorithms}
    \begin{ruledtabular}
        \begin{tabular}[c]{lllllll}
            \textbf{Algorithm} &
            \textbf{Conservation property} &
            \textbf{Deposition} &
            \textbf{Interpolation} &
            \textbf{Poisson} &
            \textbf{$E'(x)$ continuous} &
            \textbf{Field solve accuracy} \\
            & &
            \textbf{width} &
            \textbf{width} &
            \textbf{stencil width} &
            & \\
            \hline
            MC-PIC1 & Momentum & $2\Delta x$ & $2\Delta x$ & 3 & No & $O(\Delta x^2)$ \\
            EC-PIC1 & Energy (when $\Delta t \to 0$) & $2\Delta x$ & $1\Delta x$ (from edges) & 3 & No & $O(\Delta x^2)$ \\
            EC-PIC2-Lagrange & Energy (when $\Delta t \to 0$) & $3\Delta x$ & $2\Delta x$ (from edges) & 5 & No & $O(\Delta x^2)$ \\
            EC-PIC2-Standard & Energy (when $\Delta t \to 0$) & $3\Delta x$ & $2\Delta x$ (from edges) & 3 & No & $O(\Delta x^2)$ \\
            EC-PIC2-Fourth & Energy (when $\Delta t \to 0$) & $3\Delta x$ & $2\Delta x$ (from edges) & 5 & No & $O(\Delta x^4)$ \\
            CS-PIC & Energy (when $\Delta t \to 0$) & $2\Delta x$ & $2\Delta x$ & - & Yes & $O(\Delta x^4)$ \\
        \end{tabular}
    \end{ruledtabular}
    \label{tab:algos}
\end{table*}

We considered the stability of six algorithms over the parameter space $\bar{v}_t \in [0, 0.35]$ and $\bar{v}_d \in [0, 0.45]$.
\Cref{tab:algos} gives a summary of the properties of each algorithm.

\begin{figure*}
    \begin{center}
        \includegraphics{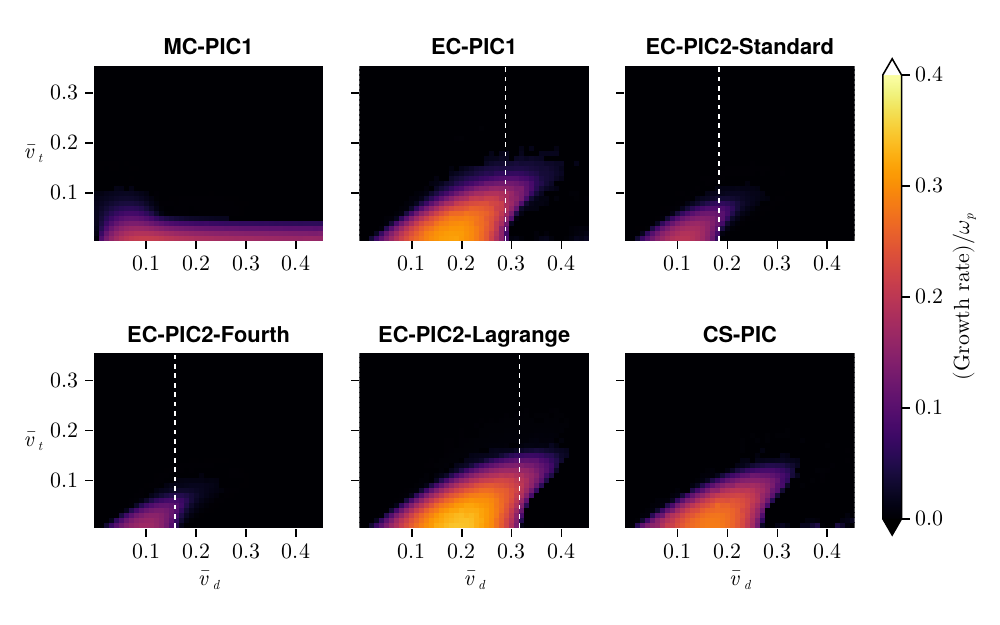}
    \end{center}
    \caption{Grid-instability growth rate as a function of the normalized drift velocity, $\bar{v}_d$, and the normalized thermal velocity, $\bar{v}_t = \lambda_D / \Delta x$, for six different PIC algorithms.
    On the basis of the results in \cref{fig:mcpic1_stationary_growth}, we estimate that our method is sensitive to growth rates down to $\gamma / \omega_p = 10^{-2}$.
    Areas of complete black indicate places where no growth rate was extracted.
    The vertical white dashed lines on the EC-PIC1 and EC-PIC2 plots indicate the predicted stability thresholds of the analytical cold-beam instability calculations presented in \cref{sec:cold_beam}.}
    \label{fig:param_space_plots}
\end{figure*}

For each algorithm, the $(\bar{v}_d, \bar{v}_t)$ parameter space was scanned with results are shown in \cref{fig:param_space_plots}.
Due to the method used to extract the growth rates, the plots show some noise, and thus some marginally unstable simulations on the periphery of the unstable region may be missed.

To stabilize a simulation, the cell size must be decreased, which moves a simulation up and to the right on the stability plot.
If the cell size is halved, the point will double in distance from the origin.
Thus, decreasing the cell size will eventually push a simulation into the stable region.

\begin{figure}
    \includegraphics{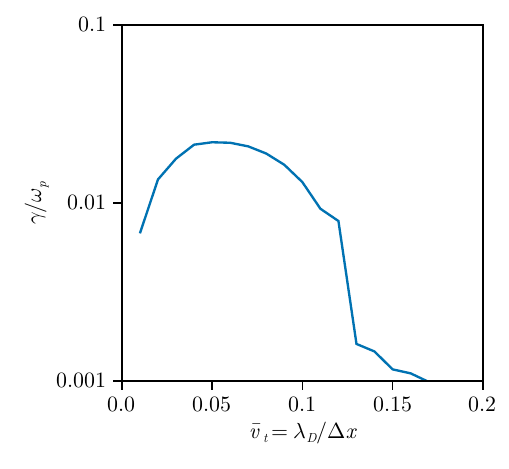} \\
    \caption{%
        Normalized grid-instability growth rates for MC-PIC1 simulating stationary ($\bar{v}_d = 0$) plasmas as a function of normalized thermal velocity, $\bar{v}_t = \lambda_D / \Delta x$.
        The line ends when the growth rate extraction procedure fails to find a nonzero growth rate because the change in thermal energy, $\mathcal{E}$, is negative for values with magnitudes larger than the fitting cutoff: $10^{-5}$.
        We estimate that the grid instability stabilizes at about $\bar{v}_t \approx 0.15$, around a factor of two lower than reported by \textcite{birdsall1980}.
    }
    \label{fig:mcpic1_stationary_growth}
\end{figure}

For the MC-PIC algorithm, it is apparent that the simulations are always unstable at low thermal velocities, regardless of the drift velocity.
In agreement with \textcite{birdsall2004}, we find stability for $\bar{v}_t \approx 0.05$ at large $\bar{v}_d \gtrsim 0.3$.

Of particular interest is the vertical axis---corresponding to an underresolved, stationary plasma---and so we show a more detailed line out of this region in \cref{fig:mcpic1_stationary_growth}.
In this regime, the grid instability saturates at a much lower $\mathcal{E}$ in the transition region.
Thus, we compute the fits for this figure in the region of the time series with a cutoff of $\mathcal{E} < 10^{-5}$.

These results indicate that, for the stationary plasma, grid instability is stabilized at about $\bar{v}_t \approx 0.15$.
This is lower by a factor of two than the \mbox{threshold} reported by \textcite{birdsall1980} of $\bar{v}_t \approx 0.3$.
This means that, for stationary plasma, it is possible to run with a cell size twice as large as reported by \textcite{birdsall1980} without experiencing grid instability.

Returning to \cref{fig:param_space_plots}, we note that all of the energy-conserving algorithms display a broadly similar trend characterized by instability when $v_t \lesssim v_d$ and $v_d \lesssim v_\text{critical}$, where $v_\text{critical}$ is an algorithm-dependent cutoff.
Red ticks below each plot indicate our analytical estimates of $v_\text{critical}$ for EC-PIC algorithms, which we derive in the next section.

We observe that EC-PIC2 has a smaller region of instability and lower instability growth rates compared to EC-PIC1, in agreement with previous work.\cite{barnes2021}
Additionally, we find that the choice of field solve stencil can significantly impact the size of the instability region; however, using a more accurate field solve does not necessarily suppress the grid instability.
An explanation for this behavior will be provided in the next section.

The CS-PIC algorithm displays a region of instability and instability growth rates on par with EC-PIC1.
We surmise that this results from cubic-spline modes that poorly replicate the intended Fourier modes, resulting in the same subgrid mode coupling that drives all grid instabilities.
Thus, it is not sufficient to ensure smoothness of the interpolated electric field.
Any algorithm that hopes to remove grid instability must eliminate the aliasing of subgrid modes.

A crucial feature of all the energy-conserving algorithms is that they are always stable when the plasma is not drifting (i.e.~$v_d = 0$).
This can be understood physically by noting that the cold-beam instability acts by transforming the bulk drift energy of the plasma into thermal energy---if there is no drift velocity then there is no energy reservoir to drive the instability.
This property makes energy-conserving algorithms an excellent choice for a wide variety of simulations of stationary plasmas.

\section{Cold-beam stability limits for EC-PIC algorithms} \label{sec:cold_beam}
It can be shown that the PIC dispersion relation is
\begin{equation}
    0 = D(k, \omega) = 1 - \frac{\omega_p^2}{K^2(k)}
    \sum_p \frac{k_p \kappa(k_p) S^2(k_p)}{(\omega - k_p v_0)^2},
\end{equation}
where $-\pi / \Delta x < k \leq \pi / \Delta x$ and where $k_p = k - 2 \pi p / \Delta x$ are the aliasing modes for a given $k$ (with integer $p$).\cite{birdsall2004,barnes2021}
The sum over $p$ occurs because the charge deposition and field interpolation couple the modes that are represented on the grid ($p = 0$) with subgrid modes of the particle density ($p \neq 0$).
The $K^2(k)$ term are the eigenvalues of the Poisson solve, and the $\kappa(k_p)$ terms are $-i$ times the Fourier representation of the gradient operator that transforms the potential to the electric field.
Note that the eigenvalues of the Poisson solve only depend on $k$ because the Poisson solve happens on the grid and so it is not influenced by the subgrid modes.
For the MC-PIC algorithm, the same is true of the finite-difference derivative of the potential, and so we have $\kappa(k_p) = \kappa(k)$.

In contrast, the energy-conserving algorithms derived in this paper compute the derivative of the potential exactly at all wavenumbers by analytically differentiating the shape function.
Thus we have $\kappa(k_p) = k_p$, which acts as a derivative on one of the factors of $S$.
Then, for the $m$th order EC-PIC algorithm, the dispersion relation is
\begin{align}
    D(k, \omega)
    &= 1 - \frac{\omega_p^2}{K^2(k)} \times \notag \\
    &\qquad\quad \sum_p \frac{k_p^2}{(\omega - k_p v_d)^2}
    \sinc^{(2m+2)} \left( \frac{k_p \Delta x}{2} \right).
\end{align}
This can be rewritten in terms of dimensionless quantities as
\begin{align}
    D(k, \omega)
    &= 1 - \left( \frac{2}{\Delta x} \right)^{(2m + 2)} \frac{\omega_p^2}{K^2(k)} \times \notag \\
    &\qquad\quad \sum_p \frac{1}{k_p^{2m}(\omega - k_p v_d)^2}
    \sin^{(2m+2)} \left( \frac{k_p \Delta x}{2} \right) \\
    &= 1 - \frac{2^{2m + 2}}{\bar{K}^2(\bar{k})} \times \notag \\
    &\qquad\quad \sum_p \frac{1}{\bar{k}_p^{2m}(\bar\omega - \bar{k}_p \bar{v}_d)^2}
    \sin^{(2m+2)} \left( \frac{\bar{k}_p}{2} \right),
\end{align}
where $\bar\omega = \omega / \omega_p$, $\bar{k} = k \Delta x$, and $\bar{K}^2(\bar{k}) = K^2(\bar{k} / \Delta x) \Delta x^2$.

The modes of the (numerical) plasma are the zeros of the dispersion relation.
The goal of this section is to compute the largest value of $\bar{v}_d$ for which the frequency of at least one mode has a positive imaginary component when $\bar{v}_t = 0$.
One way to do this---explored thoroughly by \textcite{barnes2021} for energy-conserving PIC algorithms and by \textcite{werner2025} for MC-PIC---is to scan over the range of possible $\bar{k}$ values, and numerically search for values of $\omega$ that satisfy $D(\bar{k}/\Delta x, \omega) = 0$.
Although thorough, this approach does not give much insight into the cause of the numerical instability.
Instead, we make some reasonable approximations for where the dispersion relation is likely to retain complex roots as $\bar{v}_d$ is increased.
The results of these approximate calculations show excellent agreement with the simulation results presented in the previous section, while also offering some physical insight into the mechanism of instability.

We begin by assuming that the instability will occur at $\bar{k} = \pm \pi$.
This is justified by recalling that grid instabilities result from an unphysical coupling between the grid and sub-grid modes of the particle density due to aliasing during the interpolations.
This coupling occurs because the wavenumber representation of $S$ expands beyond the range of grid wavenumbers and so a good choice of $S$ will fall off sharply beyond the Nyquist wavenumber (of course, an ideal interpolation would be zero for $|\bar{k}_p| > \pi$, but such an interpolation would not be local).
Thus, the coupling will be largest for modes just larger than the Nyquist wavenumber, and these subgrid modes will couple most strongly to grid modes just inside the Brillouin zone.
For this reason, the coupling is typically most detrimental at the Nyquist mode, $\bar{k} = \pi$ (corresponding to $\bar{k}_p = \pi - 2 \pi p = \pi q$ where $q = 1 - 2p$ is an odd integer).
Substituting this into the dispersion relation, we find
\begin{align} \label{eq:nyquist_dispersion}
    D(\pi/\Delta x, \omega)
    &= 1 - \left( \frac2\pi \right)^{2m + 2} \frac{1}{\bar{K}^2(\pi)} \times \notag \\
    &\qquad\quad \sum_{q \text{ odd}} \frac{1}{q^{2m}(\bar\omega / \pi - q \bar{v}_d)^2}
\end{align}
where we have used the fact that the sine term will become one, regardless of the value of $p$, because it is squared.

The modes of the system for $\bar{k} = \pi$ are the solutions of $D(\pi / \Delta x, \omega) = 0$.
These solutions occur when sum of the terms (each containing a pole of order two with a negative coefficient) is equal to one.
Between every pole, there will be two zeros, and for a sufficiently large drift velocity, the distance between the poles in $\bar\omega$ is much larger than the width of the poles.
In this case, each pole will be associated with two real zeros of the dispersion relation, and all of the zeros will be real.

To quantify the width of a pole, we consider the difference in $\bar\omega$ between the two zeros of an isolated pole (i.e. \cref{eq:nyquist_dispersion} with only a single term from the sum).
The zeros $\bar\omega_q$ of the isolated pole associated with term $q$ are
\begin{align}
    \bar\omega_q &= q \pi \bar{v}_d \pm 2\left( \frac2\pi \right)^{m} \frac{1}{\sqrt{\bar{K}^2(\pi)}} \frac{1}{q^{m}}
\end{align}
so that the width of the pole is
\begin{align}
    \Delta\bar\omega_q &= 4\left( \frac2\pi \right)^{m} \frac{1}{\sqrt{\bar{K}^2(\pi)}} \frac{1}{|q|^{m}}.
\end{align}

\begin{figure}
    \includegraphics{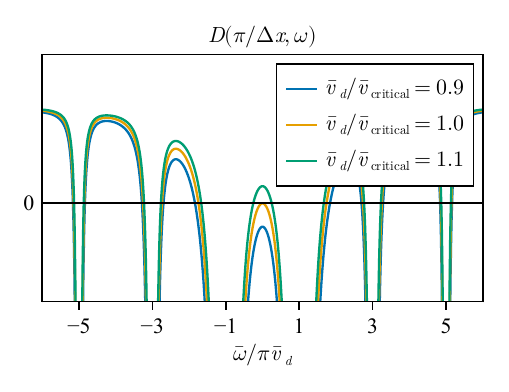}
    \caption{%
        Dispersion relation for cold-beam ($\bar{v}_t = 0$) Nyquist mode ($\bar{k} = \pi$) of EC-PIC1 at three values of $\bar{v}_d / \bar{v}_\text{critical}$.
        The resonant frequencies of the Nyquist modes are the zeros the dispersion relation, which occur between the poles.
        When $\bar{v}_d = \bar{v}_\text{critical}$, the maximum of $D(\pm \pi / \Delta x, \bar\omega)$ is equal to zero, and so all of the Nyquist mode frequencies are real.
    }\label{fig:disp_rel_plot}
\end{figure}

As $\bar{v}_d$ decreases, the pole spacing ($2 \pi \bar{v}_d$) decreases while the width of each pole remains constant.
Eventually, this causes neighboring poles to begin to overlap significantly.
When this happens, a zero of the dispersion relation can become suddenly complex as the overlap between two adjacent poles exceeds one.
The poles associated with $q = \pm 1$ are the widest, and so the region between these poles will be the first (as $\bar{v}_d$ is decreased) to exceed one.
This is shown in \cref{fig:disp_rel_plot}, where the first transition from real to complex roots occurs at $\bar\omega = 0$ when $\bar{v}_d = \bar{v}_\text{critical}$.

We can estimate the $\bar{v}_\text{critical}$ where this occurs by calculating where the width of the $q=\pm1$ poles are equal to the pole spacing.
This yields
\begin{equation}
    \bar{v}_\text{critical} \approx \frac{1}{\sqrt{\bar{K}^2(\pi)}} \left( \frac2\pi \right)^{m+1}
\end{equation}
For EC-PIC1, we have $m=1$ and the three-point Poisson solve has eigenvalues given by
\begin{equation}
    \bar{K}^2(\bar{k}) = \bar{k}^2 \sinc\left(\frac{\bar{k}}{2}\right),
\end{equation}
so that
\begin{equation}
    \bar{v}_\text{critical} \approx \frac{1}{\sqrt{4}} \left( \frac2\pi \right)^{2} = \frac{2}{\pi^2} \approx 0.203.
\end{equation}

However, we can do much better than this order of magnitude estimate by noting that the poles are symmetric about $\bar\omega=0$, and so the first pair of zeros to develop complex components will both be $\bar\omega = 0$ just before they become complex.
Substituting $\bar\omega=0$ into the dispersion relation yields
\begin{align}
    D(\pi, 0)
    &= 1 - \left( \frac2\pi \right)^{2m + 2} \frac{1}{\bar{K}^2(\pi)} \frac{1}{\bar{v}_d^2}
    \sum_{q \text{ odd}} \frac{1}{q^{2m + 2}}.
\end{align}
For $m = 1$, we have
\begin{equation}
    \sum_{q \text{ odd}} \frac{1}{q^{4}} = \frac{\pi^4}{48},
\end{equation}
and so
\begin{align}
    D(\pi, 0)
    &= 1 - \frac{1}{12} \frac{4}{\bar{K}^2(\pi)} \frac{1}{\bar{v}_d^2}.
\end{align}
Therefore, the EC-PIC1 algorithm will be stable for
\begin{align}
    \bar{v}_d > = \bar{v}_\text{critical} = \sqrt\frac{1}{12} \approx 0.288.
\end{align}
This differs from the stability threshold reported in \textcite{birdsall1975} which claims $\bar{v}_\text{critical} \approx 1 / \pi$, although the numerical values are remarkable similar.\cite{birdsall1980,birdsall2004}
Additionally, this critical drift velocity agrees very well with the simulation results presented in \cref{fig:param_space_plots}.

\begin{table}
    \begin{ruledtabular}
        \begin{tabular}[c]{lll}
            \multicolumn{1}{c}{\textbf{Field Solve}} & 
            \multicolumn{1}{c}{\textbf{Eigenvalues ($\bar{K}^2(\bar{k})$)}} & 
            \multicolumn{1}{c}{\textbf{$v_\text{critical}$}} \\
            \hline
            Three-point finite-difference &
            $ \bar{k}^2 \sinc^2\left(\frac{\bar{k}}{2}\right) $ &
            $ \sqrt\frac{1}{30} \approx 0.183 $ \\
            Five-point finite-difference &
            $ \bar{k}^2 \sinc^2\left(\frac{\bar{k}}{2}\right) \frac{7 - \cos \bar{k}}{6} $ &
            $ \sqrt{\frac{1}{40}} \approx 0.158 $ \\
            Lagrangian-derived field solve &
            $ \bar{k}^2 \sinc^2\left(\frac{\bar{k}}{2}\right) \frac{2 + \cos \bar{k}}{3} $ &
            $ \sqrt\frac{1}{10} \approx 0.316 $ \\
        \end{tabular}
    \end{ruledtabular}
    \caption{Critical drift velocities for the EC-PIC2 algorithm with a variety of Poisson solves.}
    \label{tab:ecpic2_vcrit}
\end{table}

The same analysis can be repeated for EC-PIC2 (i.e. $m=2$) which yields
\begin{align}
    D(\pi, 0)
    &= 1 - \frac{2^6}{\pi^6} \frac{1}{\bar{K}^2(\pi)} \frac{1}{\bar{v}_d^2} \frac{\pi^6}{480} \\
    &= 1 - \frac{2}{15} \frac{1}{\bar{K}^2(\pi)} \frac{1}{\bar{v}_d^2}.
\end{align}
It is apparent that different field solves will have different eigenvalues and different critical velocities.
For the three field solves considered in the previous section, the Nyquist eigenvalue and corresponding critical velocity are shown in \cref{tab:ecpic2_vcrit}.
Once again, these analytical predictions agree well with the growth rates computed from simulations in the previous section (see \cref{fig:param_space_plots}).

\section{Concluding remarks} \label{sec:conclude}
If the Debye length of a plasma is not sufficiently well resolved---for example because doing so would be too computationally expensive---then the plasma will heat until the Debye length becomes resolved.
This grid instability has the potential to destroy the integrity of a simulation.
Although it is relatively easy to detect in these simple simulations when the Debye length is sufficiently underresolved---monitoring plasma temperature is sufficient---it can become extremely challenging to determine if heating in more complex simulations is due to physical or numerical effects.

Requiring resolution of the Debye length means that simulations of many physical problems of interest are infeasible or borderline infeasible due to the scale disparity between the device geometry and the Debye length.\cite{ryabinkin2021,kanarik2020,kanarik2023}
Thus careful characterization of grid instability is necessary to determine the minimum resolution required for stability.
For the standard PIC method (MC-PIC), this paper confirms the existence of a stability threshold in Debye length resolution at large drift velocity (instability for $\bar{v}_t = \lambda_D / \Delta x \lesssim 0.05$ when $\bar{v}_d \gtrsim 0.15$).\cite{birdsall2004}
Additionally, we report new measurement of grid instability for plasmas at very low drift velocities (i.e. $\bar{v}_d \lesssim 0.01$), which indicate that grid instability occurs at a lower threshold---$\bar{v}_t \approx 0.15$---than previously reported.

There exists a now fifty-year-old algorithm (EC-PIC) that is capable of circumventing some of these restrictions.\cite{lewis1970,langdon1973}
Recent papers have sought to bring more attention to this method and to establish stability limits\cite{evstatiev2013,barnes2021,powis2023}
We have confirmed the instability thresholds reported by \textcite{barnes2021}, as well measured instability thresholds for two different field solves in the EC-PIC2 algorithm.
All algorithms display a broadly similar instability behavior with instability when $\bar{v}_t \lesssim \bar{v}_d$ and when $\bar{v}_d \lesssim \bar{v}_\text{critical}$.
Additionally, we have derived analytical estimates for $\bar{v}_\text{critical}$ for several of EC-PIC variants, all of which show excellent agreement with our simulation results.
As a result of this analysis, we have shown that use of wider shape functions decreases $\bar{v}_\text{critical}$ (that is, reduces the region of the $(\bar{v}_d, \bar{v}_t)$ parameter space which is unstable).
We have also shown that a more accurate field solve reduces $\bar{v}_\text{critical}$, but the use of an accurate field solve cannot totally eliminate the grid instability.
Both of these analytical results are reflected in our simulation data.
We emphasize previous work which finds that energy-conserving algorithms are suitable for simulating stationary plasmas.\cite{barnes2021,powis2023}

Finally, we have presented a new PIC algorithm that uses cubic interpolating splines to represent fields (CS-PIC).
The splines approximate a truncated Fourier basis for the potential, which has been shown to be unconditionally stable to grid instability.\cite{evstatiev2013,mitchell2019}
We have shown that the interpolations to and from the cubic-spline basis can be efficiently computed by having two charge density fields, and two electric potential fields and performing two FFTs in the field solve.
This novel technique for calculating cubic splines avoids a linear solve to compute the second derivative of the splines at the nodes.

The interpolated electric field in CS-PIC has a continuous first derivative.
However, we find that the stability behavior of CS-PIC is only a modest improvement over EC-PIC1, which indicates that the use of a smoother electric does not significantly impact the growth of the grid instability.
The CS-PIC method can be readily extended to ensure higher-order differentiability of the interpolated electric field, becoming a sort of unequally-spaced FFT as the number of deposition and interpolation fields are increased.
In the limit of some extended CS-PIC exactly approximates PIF, it should conserve both energy and momentum; however we have not given detailed thought to the computational cost of this approach compared to PIF.

Due to the complexity of CS-PIC relative to EC-PIC, we recommend that readers first evaluate whether EC-PIC1 or EC-PIC2 is sufficient to simulate the physical problem in question.

\begin{acknowledgments}
    This work has been supported by the Air Force Office of Scientific Research, grant number AFOSR FA9550-18-1-0436, and the National Science Foundation, grant numbers NSF (PHY) 2206647 and NSF (PHY) 2206904.
\end{acknowledgments}

\section*{Author Declarations}
The authors have no conflicts to disclose.

\section*{Data Availability Statement}
The code used to produce these results is available at \url{https://github.com/adamslc/ECPIC-paper-code}, and archived at \url{https://doi.org/10.5281/zenodo.15022285}.
The resulting data is archived at \url{https://doi.org/10.5281/zenodo.15022224}.

\appendix*
\section{Orthogonality of cubic-interpolating-spline approximations to Fourier modes}
The cubic-interpolating-spline approximation to a Fourier mode can be written as
\begin{align} \label{eq:cs_fourier}
    e_k(x)
    &= \sum_{n=0}^{N} e^{ikX_n} \left( w_1(\xi_n) + \hat{C}_k \Delta x^2 \, w_3(\xi_n) \right) \\
    &= \sum_{n=0}^{N} e^{ikX_n} g_n^{(k)}(x),
\end{align}
where $g_n^{(k)}(x)$ is the cubic interpolating polynomial for wavenumber $k$ at node $n$.
Recall that each polynomial has been constructed so that it is only nonzero on the interval $(X_{n-1}, X_{n+1})$, and that $g_n^{(k)}(x) = g_0^{(k)}(x - n \Delta x)$.
It then follows that
\begin{align}
    \langle e_k | e_{\ell} \rangle
    &= \sum_{n=0}^{N} \sum_{m=n-1}^{n+1} e^{ikX_n} e^{-i\ell X_m} \int_0^L \dd{x} g_n^{(k)}(x) g_m^{(\ell)}(x) \\
    &= \sum_{n=0}^{N} e^{i(k - \ell)X_n} \sum_{r=-1}^{1} e^{i\ell r \Delta x} \int_0^L \dd{x} g_0^{(k)}(x) g_r^{(\ell)}(x) \\
    &= \left(\sum_{r=-1}^{1} e^{i\ell r \Delta x} \int_0^L \dd{x} g_0^{(k)}(x) g_r^{(\ell)}(x)\right) \sum_{n=0}^{N} e^{i(k - \ell)X_n} \\
    &= \left(\sum_{r=-1}^{1} e^{i\ell r \Delta x} \int_0^L \dd{x} g_0^{(k)}(x) g_r^{(\ell)}(x)\right) N \delta_{k\ell}.
\end{align}
Thus, the basis functions $\{e_k\}$ are orthogonal.
A similar argument proves the orthogonality of the $\{e'_k\}$ functions.
It follows that the CS-PIC field solve will be exactly diagonal in Fourier space (i.e. nonzero only for $k = \ell$).

For the diagonal elements, the inner products can be evaluated exactly.
We first note that the integral is identical for $r = \pm 1$, and so
\begin{align}
    \langle e_k | e_k \rangle
    &= N \left[ \int \dd{x} \left(g_0^{(k)}(x)\right)^2
    + 2 \cos(k \Delta x) \int \dd{x} g_0^{(k)}(x) g_1^{(k)}(x) \right]
\end{align}
Then it can be shown that
\begin{equation}
    \langle e_k | e_k \rangle = L \frac{272 + 297c + 60 c^2 + c^3}{70 (2 + c)^2}
\end{equation}
where $c = \cos(k \Delta x)$.
In the long wavelength limit, $c = 1$ and the inner product reduces to $\langle e_k | e_k \rangle = L$ as expected.
A similar argument yields
\begin{align}
    \langle e'_k | e'_k \rangle
    &= N \left[ \int \dd{x} \left(g\prime_0^{(k)}(x)\right)^2
    + 2 \cos(k \Delta x) \int \dd{x} g\prime_0^{(k)}(x) g\prime_1^{(k)}(x) \right] \\
    &= \frac{L}{\Delta x^2} \frac{48 - 9c - 36 c^2 - 3 c^3}{5 (2 + c)^2}.
\end{align}
This inner product reduces to $\langle e'_k | e'_k \rangle = k^2 L$ in the small $k$ limit.

\nocite{*}
\bibliography{zotero}

\end{document}